\documentclass[10pt,letterpaper]{article}
\usepackage[top=0.85in,left=2.75in,footskip=0.75in]{geometry}

\usepackage{changepage}
\usepackage{float}
\usepackage[utf8]{inputenc}

\usepackage{textcomp,marvosym}

\usepackage{fixltx2e}

\usepackage{amsmath,amssymb}

\usepackage{cite}

\usepackage{nameref,hyperref}

\usepackage{microtype}
\DisableLigatures[f]{encoding = *, family = * }

\usepackage{rotating}

\raggedright
\usepackage[aboveskip=1pt,labelfont=bf,labelsep=period,justification=raggedright,singlelinecheck=off]{caption}

\makeatletter
\renewcommand{\@biblabel}[1]{\quad#1.}
\makeatother

\date{}

\usepackage{lastpage,fancyhdr,graphicx}
\usepackage{epstopdf}
\fancyheadoffset[L]{2.25in}
\fancyfootoffset[L]{2.25in}

\begin{document}
\vspace*{0.35in}

\begin{flushleft}
{\Large
\textbf\newline{Developing and Testing a Bayesian Analysis of Fluorescence Lifetime Measurements} 
}
\newline
\\
Bryan Kaye\textsuperscript{1,2,\Yinyang,*},
Peter J. Foster\textsuperscript{1,2,\Yinyang},
Tae Yeon Yoo\textsuperscript{1,2,\Yinyang},
Daniel J. Needleman\textsuperscript{1,2,3}
\\
\bigskip
\bf{1} John A. Paulson School of Engineering and Applied Sciences, Harvard University, Cambridge, MA, United States
\\
\bf{2} FAS Center for Systems Biology, Harvard University, Cambridge, MA, United States
\\
\bf{3} Department of Molecular and Cellular Biology, Harvard University, Cambridge, MA, United States
\\
\bigskip

%
%
\Yinyang  These authors contributed equally to this work.





*bryankaye@fas.harvard.edu

\end{flushleft}
\section*{Abstract}
FRET measurements can provide dynamic spatial information on length scales smaller than the diffraction limit of light. Several methods exist to measure FRET between fluorophores, including Fluorescence Lifetime Imaging Microscopy (FLIM), which relies on the reduction of fluorescence lifetime when a fluorophore is undergoing FRET. FLIM measurements take the form of histograms of photon arrival times, containing contributions from a mixed population of fluorophores both undergoing and not undergoing FRET, with the measured distribution being a mixture of exponentials of different lifetimes. Here, we present an analysis method based on Bayesian inference that rigorously takes into account several experimental complications. We test the precision and accuracy of our analysis on controlled experimental data and verify that we can faithfully extract model parameters, both in the low-photon and low-fraction regimes.

\section*{Introduction}
F\"{o}rster resonance energy transfer, or FRET, is a fluorescence technique commonly used to access spatial information on length scales smaller than the diffraction limit of light \cite{Roy:2008p1081}. In standard fluorescence, illuminating light is used to excite a fluorophore into a higher energy state, and the fluorophore subsequently relaxes into its ground state either by emitting a photon or through a non-radiative decay pathway. If another fluorophore is near, typically within $\approx$ 10 nm, the two fluorophores can interact through dipole-dipole interactions termed FRET. FRET confers an additional decay path where the excited florophore, termed the donor, can transfer its energy to the nearby, unexcited fluorophore, termed the acceptor, which can then release the energy as a photon or through non-radiative decay. As the emission spectra of commonly used donor and acceptor pairs are spectrally distinct, one common method of measuring the the average FRET efficiency is to compare the relative intensities collected from the two channels. However, this method has drawbacks including spectral bleed-through and a sensitivity to changes in fluorophore concentration and excitation light intensity \cite{Wallrabe:2005bk}.

As an alternative to using fluorescence intensity to quantify FRET, fluorescence lifetime imaging microscopy, or FLIM, can be used \cite{Stachowiak:2012gg,Peter:2005eg,Yoo:2016cu,Hinde:2013eo}. In time-domain FLIM, a narrow pulse of light is used to excite fluorophores into an excited state. Fluorophores that decay from their excited states can do so by releasing a photon. A subset of the released photons are detected, and for each detected photon, the arrival time is measured relative to the excitation pulse.  The amount of time fluorophores spend in their excited state depends on the number of decay paths available. Donor fluorophores are chosen such that when they decay from their excited states, they do so at a constant rate, leading to photon emission time distributions that are exponential with a single characteristic decay time. This characteristic decay time is known as the fluorescence lifetime and is typically on the order of nanosconds. When donor fluorophores are undergoing  FRET, they will spend, on average, a shorter amount of time in their excited states, leading to a reduced lifetime. In a sample where only a fraction of donor fluorophores are undergoing FRET, the photon emission time distribution will be the sum of two exponentials with different lifetimes. By comparing the amplitudes of these two exponentials, the relative fraction of donors undergoing FRET can be measured. In practice, additional complications are present, including photons collected from spurious background and time delays introduced by the collection system itself. These effects must be accounted for to infer the relative amplitudes and lifetimes of the emitted photon distributions from the measured photon arrival time histograms. Several approaches have been used in order to estimate these parameters, including least-squares fitting \cite{Chang:2007je}, phasor methods \cite{Stringari:2011ju,Colyer:2012jba,Chen:2015ev}, and Bayesian approaches \cite{Rowley:2011kj}, each with their own advantages and disadvantages. 

Here we utilize and extend the Bayesian approach previously described \cite{Rowley:2011kj} to take into account additional experimental factors and test the performance of our method using experimental data. 

\section*{Materials and Methods}

\subsection*{Bayesian Framework}

Our framework is based on a previously described Bayesian analysis approach for measuring lifetimes from FLIM data \cite{Rowley:2011kj}. Bayes' Law states that given a set of data, $t$, and a set of model parameters $\theta$, then, 
\begin{equation}\label{eq:Bayes' Law} p(\theta|t) \propto p(t|\theta) \times p(\theta)\end{equation}

\noindent where $p(\theta|t)$, the probability of the model parameters given the measured data, is referred to as the posterior distribution, $ p(t|\theta) $ is referred to as the likelihood function, and $p(\theta)$ is referred to as the prior distribution. The aim of Bayesian inference approaches is to find the posterior distribution for the given model and data, and hence what the probability is for each possible set of model parameters $\theta$. 

In time-domain FLIM measurements, a narrow laser pulse is used to excite fluorophores in the sample, and the arrival times of photons emitted from the fluorophores are recorded. Fluorophores undergoing FRET will have a shorter florescence lifetime compared with fluorophores not undergoing FRET. When only a fraction of fluorophores in the sample are undergoing FRET, the resulting distribution of photon emission will be a sum of exponentials, where each exponential has a different lifetime, and each exponential is weighted by the number of photons collected from the respective source.  In addition, there exists a constant background of photons due to noise in the detector and stray light, taken to be from a uniform distribution. In the following, we consider photons from each of these sources separately and construct the likelihood function as follows,

\begin{equation}\label{eq:Continuous Likelihood}  p(t|\theta) = f_S\times p_S(t|f_S,\tau_S) + f_{L} \times p_L(t|f_{L},\tau_L) + f_B \times p_B(t|f_B)   \end{equation}

\noindent where t is the arrival time of a photon relative to the excitation pulse, $\tau_S$ and $\tau_L$ are respectively the short and long fluorescence lifetimes, $f_S$ and $f_L$ are the fractions of photons from the short and long lifetime distributions respectively, $f_B$ is the fraction of photons from the uniform background given by $f_B = (1-f_S -f_L)$. Here $p_i(t|f_i)$ is the probability of the photon arriving at time t given that the photon originates from fraction $f_i$.

Equation \ref{eq:Continuous Likelihood} represents the likelihood model when time is taken to be continuous. However, in practice, photon arrival times collected with TCSPC are discretized into bins, and this discretization must be taken into account. If the bins are numbered sequentially and of width $\Delta t$, such that $b_i$ represents the bin containing photons with arrival time, $(i-1)\Delta t \leq t \leq i \Delta t$, then the likelihood function becomes,

\begin{align}\label{discreet_likelihood} 
p(t|\theta) = \prod_{i=1}^N \Big[ f_S\times p_S(t\in b_i|f_S,\tau_S) + f_{L} \times p_L(t\in b_i|f_{L},\tau_L) +  \nonumber \\ f_B \times p_B(t\in b_i |f_B)\Big]^{P_{i}} \end{align}

\noindent Thus, Eqn. \ref{discreet_likelihood} serves as the discrete form of the likelihood function, Eqn. \ref{eq:Continuous Likelihood}.

\subsection*{Instrument Response Function}
One complexity in experimental TCSPC measurements is that a delay is introduced to photon arrival times, termed the Instrument Response Function (IRF). In order to account for this effect, the IRF was experimentally measured (see FLIM Measurements). The measured IRF is then convolved with the idealized probability density functions for the exponential distributions in order to construct the likelihood function. Taking this effect into account leads to, 

\begin{align}\label{IRF} p_j(t\in b_i|f_j,\tau_j) = p_{em,j}(t \in b_i|f_j,\tau_j)\otimes IRF(t) \nonumber \\
\end{align}

\noindent where $p_{em,j}$ is the idealized exponential distribution, taken to be $\propto e^{-t/\tau_j}$, where $j \in \{S,L\}$ is an index labeling the exponential distribution and $IRF(t)$ is the experimentally measured instrument response function.

\subsection*{Posterior Distribution}
Using Eqn. \ref{IRF} in Eqn. \ref{discreet_likelihood} leads to the final form of our likelihood function,

\begin{align}\label{final_likelihood} 
p(t|\theta) = \prod_{i=1}^N \Big[ f_S\times p_{em,S}(t\in b_i|f_S,\tau_S)\otimes IRF(t)  \nonumber \\ + f_{L} \times p_{em,L}(t\in b_i|f_{L},\tau_L) \otimes IRF(t) \nonumber \\ + f_B \times p_B(t\in b_i |f_B)\Big]^{P_{i}} \end{align}

\noindent For comparison with experiments using control dyes where the lifetimes of the two molecules are well characterized, we choose a prior distribution such that the distribution is uniform for the fractions in the domain $f_j \in [0,1]$,  and $\tau_S$ and $\tau_L$ are set to the measured values for Coumarin 153 and Erythrosin B respectively. With this choice of prior, Eqn. \ref{eq:Bayes' Law} becomes,
\begin{equation}\label{eq:Bayes' Law no prior} p(\theta|t) \propto p(t|\theta) \end{equation}
\noindent and hence our posterior distribution is proportional to our likelihood function in the constrained parameter space. To build the posterior distribution, parameter space is searched by evaluating the likelihood function on a grid of uniform spacing. Alternatively, parameter space can be searched stochastically using the Markov chain Monte Carlo method, yielding equivalent results (Fig. \ref{fig:SFigure4}).

\subsection*{Effects of Periodic Excitation}
For a single exponential decay, the probability of measuring a photon at time t, given a decay lifetime, $\tau$, is given by,
\begin{equation}\label{eq:single pulse} p_{em} (t|{\tau}) \propto e^{-\frac{t}{\tau}}  \end{equation}
Where $\tau$ is the lifetime of the fluorophore. In practice, many sequential excitation pulses are used, and it's possible that a fluorophore excited by a given pulse doesn't emit a photon until after a future pulse. Taking this effect into account for a single exponential decay leads to \cite{Rowley:2011kj},
\begin{equation}\label{eq:multi pulse} p_{em}(t|{\tau},T) \propto \sum_{k=0}^{\infty}e^{-\frac{t+kT}{\tau}}   \end{equation}
\noindent where $T$ is the excitation pulse period and $k$ is an index counting previous pulses. The sum is a geometric series, which converges to, 
\begin{equation}\label{eq:geometric series} p_{em}(t|{\tau},T) \propto  \frac{1}{1-e^{\frac{-T}{\tau}}} e^{\frac{-t}{\tau}}  \end{equation}
Thus, accounting for periodic excitations leads to a prefactor $\frac{1}{1-e^{\frac{-T}{\tau}}}$, which for a given $T$ and $\tau$ is constant. As we treat exponentials from populations with short and long lifetimes separately, this factor can safely be absorbed into the normalization constant, leaving the probability distribution unchanged.



\subsection*{FLIM Measurements}\label{sec:FLIM Measurements} FLIM measurements were carried out on a Nikon Eclipse Ti microscope using two-photon excitation from a Ti:sapphire pulsed laser (Mai-Tai, Spectra-Physics, 865 nm or 950 nm wavelength, 80 MHz repetition rate, $\approx$ 70 fs pulse width), a commercial scanning system (DCS-120, Becker \& Hickl), and hybrid detectors (HPM-100-40, Becker \& Hickl). The excitation laser was collimated by a telescope assembly to avoid power loss at the XY galvanometric mirror scanner and to fully utilize the numerical aperture of a water-immersion objective (CFI Apo 40x WI, NA 1.25, Nikon). Fluorescence was imaged with a non-descanned detection scheme with a dichroic mirror (705 LP, Semrock) that was used to allow the excitation laser beam to excite the sample while allowing fluorescent light to pass into the detector path. A short-pass filter was used to further block the excitation laser beam (720 SP, Semrock) followed by an emission filter appropriate for Coumarin and Erythrosin B (550/88nm BP, Semrock, or 552/27nm BP, Semrock). A Becker \& Hickl Simple-Tau 150 FLIM system was used for time correlated single photon counting \cite{Becker:2010}. The instrument response function was acquired using second harmonic generation of a urea crystal \cite{Becker:2010}.

For the data shown in Figures 1 and 3, the \emph{TAC range} was set to $7\times10^{-7}$ with a \emph{Gain} of 5, corresponding to a 14 ns maximum arrival time. The \emph{TAC offset} was set to 6.27\%. The TAC \emph{limit high} and \emph{limit low} were set to 5.88\% and 77.25\%, respectively, resulting in a 10 ns recording interval. Erythrosin B and Coumarin 153 samples were prepared at 10 mM and 15 mM, respectively. Lifetimes were measured and fixed at values of 3.921 ns and 0.453 ns for Coumarin 153 and Erythrosin B respectively.

For the data recorded shown in Figures 2, the \emph{TAC range} was set to  $5\times10^{-8}$ with a \emph{Gain} of 5,  corresponding to a 10 ns maximum arrival time. The TAC \emph{limit high} and \emph{limit low} were set to 95.29\% and 5.88\%, respectively, resulting in a 10 ns recording interval. Illumination intensity was set such that $\approx2.5 \times 10^{5}$ photons per second were recorded at the photon detector. Lifetimes were measured and fixed at values of 4.03 ns and 0.48 ns for Coumarin 153 and Erythrosin B respectively.

\subsection*{Software Implementation}
All algorithms were implemented in MATLAB. The code used is freely available on Github at https://github.com/bryankaye1/bayesian-analysis-of-fluorescent-lifetime-data. Posterior distributions were generated by evaluating the likelihood function in a grid space of parameter values and were marginalized before estimation of the mode and mean for each parameter. 

\begin{figure}[h*]

\includegraphics[scale=0.75]{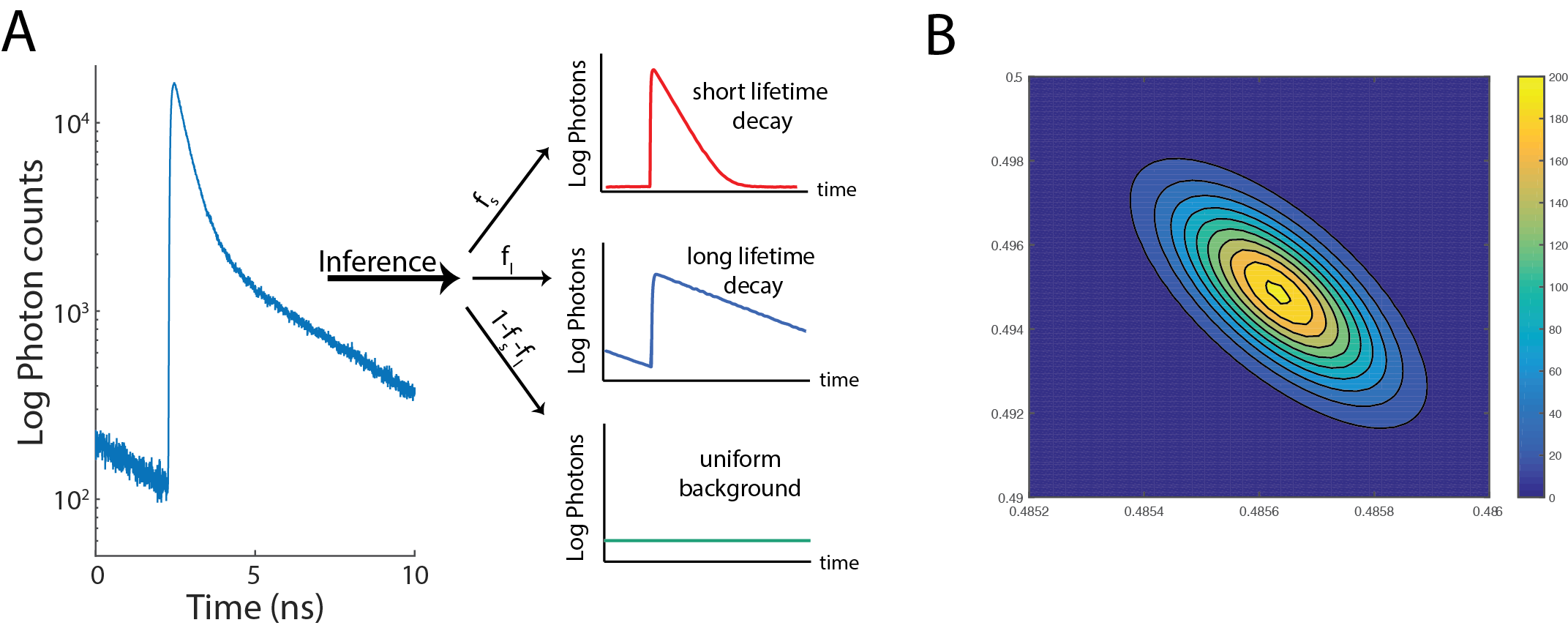}
\caption{{\bf Photon arrival-time histograms are composed of the sum of two exponential distributions}
(A) Photon arrival histogram composed of two exponential distributions, with a short-lifetime fraction $f_S$, a long-lifetime fraction $f_L$, and a background fraction $f_B=(1-f_S-f_L)$ (B) Inferred posterior distribution generated from data in Fig 1A.}
 \label{fig:Figure1}
 \end{figure}
 
\section*{Results}
In a sample where only a subset of fluorophores are undergoing FRET, photon emission distributions take the form of a biexponential distribution, with some fraction of the distribution consisting of photons from a short-lifetime exponential, another fraction consisting of photons from a long-lifetime exponential, and some fraction coming from a spurrious background distribution. The goal of FLIM analysis is to infer the relative weights of these distributions, along with the lifetimes of the two exponential distributions, from the measured histogram of photon arrival times (Fig. \ref{fig:Figure1}A). Here we apply an analysis based on Bayesian inference in order to infer the most likely set of parameters from experimentally measured data. The output of our algorithm is a posterior distribution, which gives the relative probability of measuring a given set of parameters (Fig. \ref{fig:Figure1}B). To characterize our approach, we test our analysis in both the low-photon and low-fraction regimes, representing two extremes where data may be collected.

\begin{figure}[h] 
\centering
\includegraphics[scale=0.65]{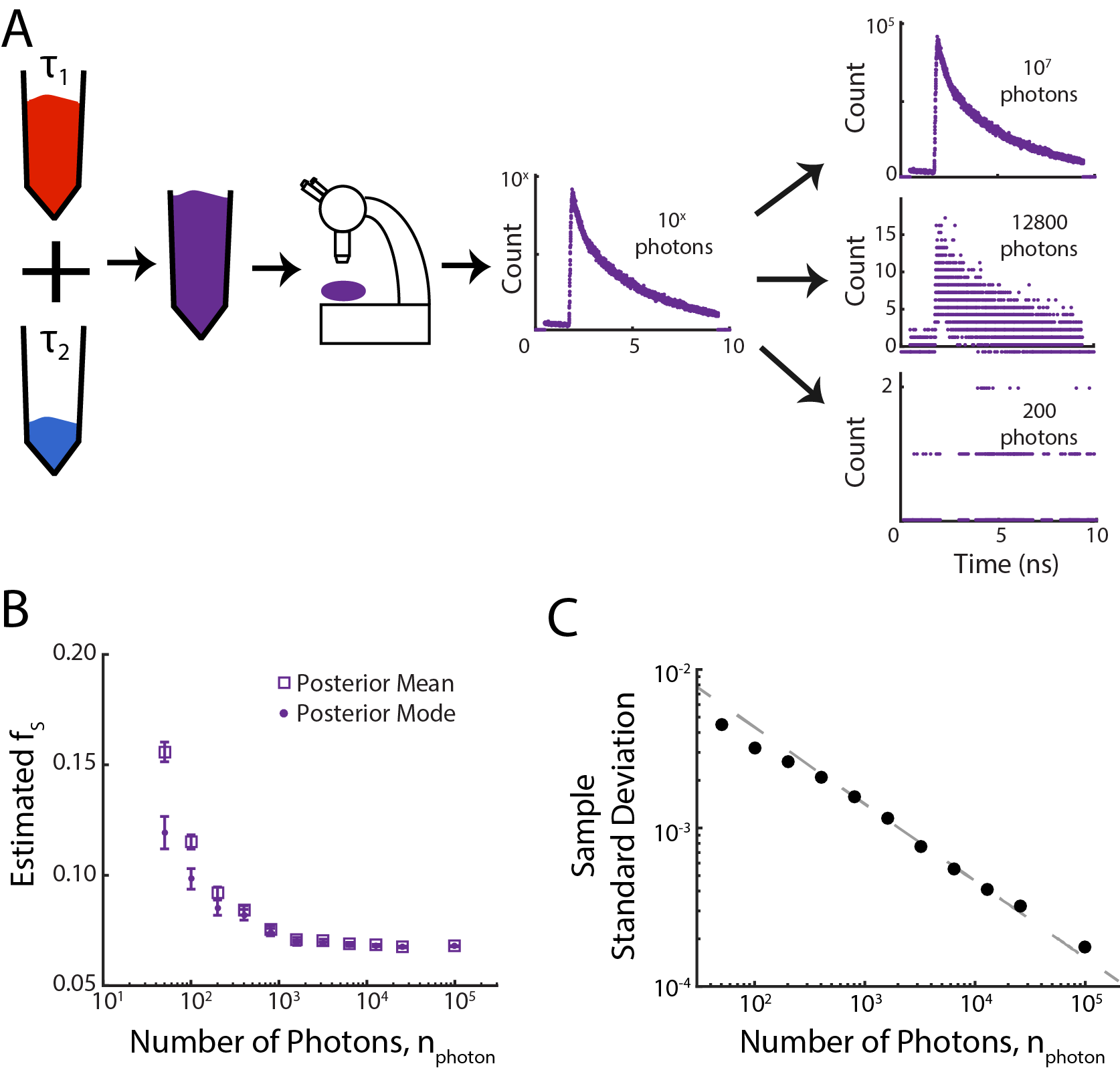}
\caption{{\bf Low-Photon Regime}
(A) Control dyes having known long (Coumarin 153) and short (Erythrosin B) lifetimes were mixed at a fixed ratio.  From the measured master curve of photon arrival times, a variable number of photons are randomly sampled, generating histograms with a variable number of photons. (B) Bias in the estimated short-life photon fraction, $f_S$, decreases with increasing photon number. Data points represent the average of the posterior mean (squares) or mode (circles) for 300 independent samplings for each photon count. Error bars are s.e.m. (C)  Black circles: measured sample standard deviations from data in Fig 2B averaged across the 300 independent samplings. The sample standard deviation decreases approximately as $\sqrt{n_{photon}}$. Power law fit to $a \times x^b$  for all but the four lowest values of $n_{photon}$ shown in gray, with $a = 0.04 \pm 0.01$ and $b=-0.48 \pm 0.04$ (95\% confidence interval) }
\label{fig:Figure2}
\end{figure}

\subsection*{Low-Photon Regime}
While the biexponential nature of FLIM histograms is apparent when the histogram is constructed using a large number of photons (Fig. \ref{fig:Figure1}A), the histogram's underlying distribution is less obvious when the photon count is low (Fig. \ref{fig:Figure2}A). Consequently, in this regime it can be difficult to extract accurate estimates of the fraction of short-lifetime photons though methods that rely on histogram fitting. This low-photon regime is relevant in many applications of FLIM, due to the fundamental tradeoff between the number of photons collected and both the spatial-temporal precision of the measurement and the light dose received by the sample. Thus, methods that can improve the precision and accuracy of parameter estimation in the low-photon count regime could potentially lead to a practical increase in spatial-temporal resolution and lower light doses. 

In order to test the accuracy and sensitivity of our analysis, fluorescence lifetime measurements were taken using Erythrosin B and Coumarin 153, two reference dyes with well characterized lifetimes of 0.47 $\pm$ 0.02 ns and 4.3 $\pm$ 0.2 ns respectively \cite{Boens:2007da}. These dyes were mixed at a fixed ratio, and fluorescence lifetime measurements were taken (Fig. \ref{fig:Figure2}A, Materials and Methods) in order to generate a master list of photon arrival times. A fixed number of photons were randomly sampled from the master list in order to construct a histogram of photon arrival times, and analyzed to infer an estimate of the fraction of short-lifetime fluorophores, $f_S$, taken as either the mean or the mode of the posterior distribution. This process was repeated 300 times in order to produce an error estimate for each given photon count, and was repeated for total photon counts spanning $\approx$ 3 orders of magnitude (Fig \ref{fig:Figure2}B). 

We find good agreement between the estimates of the fraction of short-lifetime photons for total photons counts larger than $\approx$ 200 photons, using either the posterior mean or posterior mode as a fraction estimate (Fig. \ref{fig:Figure2}B). Slight discrepancies between estimates using the posterior mean and posterior mode are apparent due to truncation and the fact that the posterior distribution is skewed (Fig. \ref{fig:Figure1}B), and thus in general the mode and the mean of the distribution are not equal. As a measure of the error in our parameter estimation, we compute the standard deviation of the estimates from the 300 numerical replicates (Fig. \ref{fig:Figure2}C) for each photon count. Fitting a power law to all data points except for the four smallest photon counts yields an exponent of $-0.48 \pm 0.04$ (95\% confidence interval), consistent with the exponent of $-0.5$ predicted from the central limit theorem in the limit of high $n_{photon}$ .

\subsection*{Low-Fraction Regime}
\begin{figure}[H]
\centering
\includegraphics[scale=0.5]{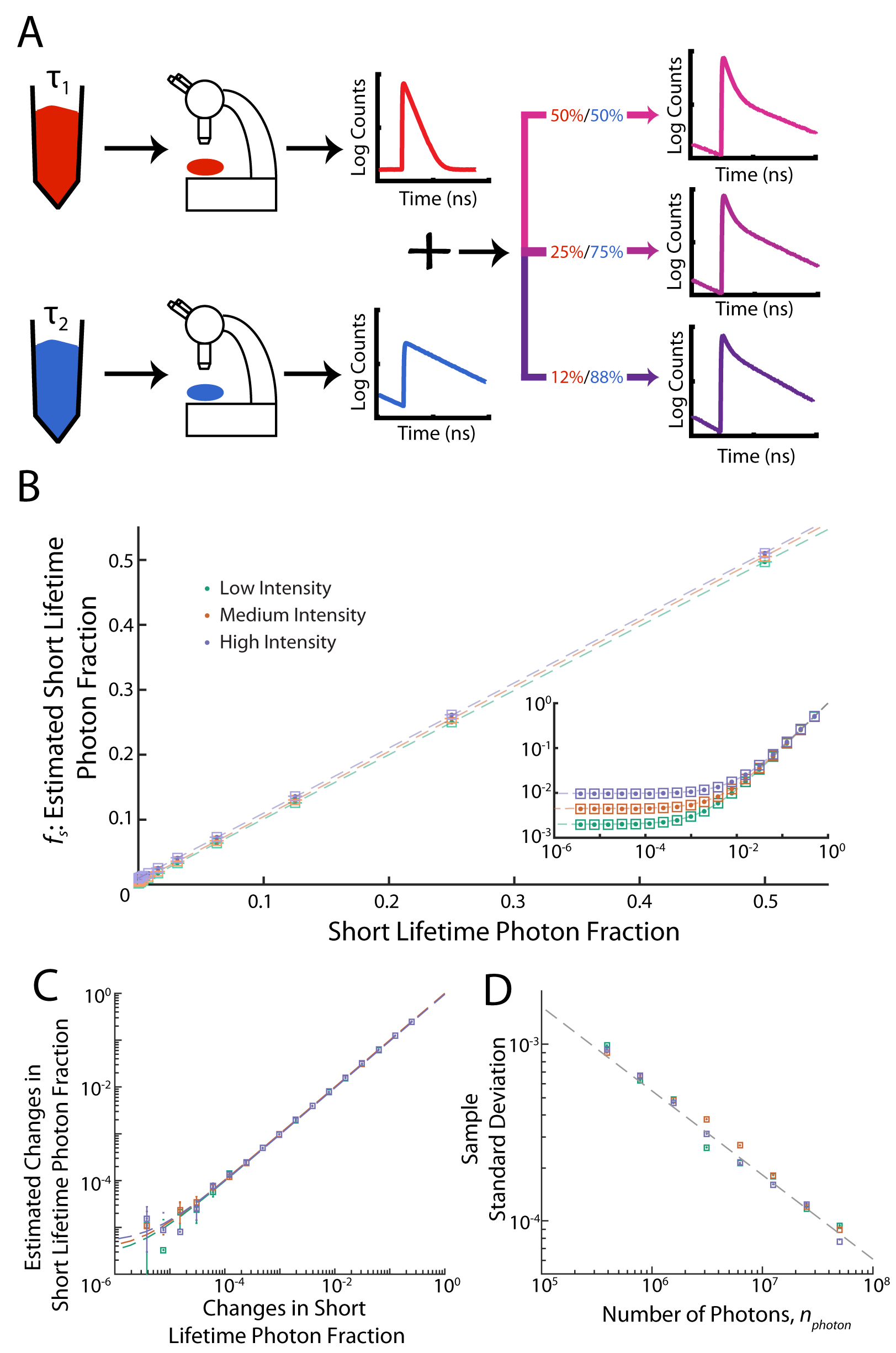}

\caption{{\bf Low-fraction regime}
(A) Samples of dyes with short-lifetime (Erythrosin B) and long-lifetime (Coumarin 153) were prepared, and fluorescence lifetime measurements were collected for each dye separately, leading to separate master photon histograms. Test histograms were constructed by randomly sampling a fixed number of photons, with varying fractions being drawn from the master lists of short-lifetime and long-lifetime photons. These histograms were then analyzed in order to estimate the fraction of short-lifetime photons, $f_S$. (B) The estimated short-lifetime fraction, $f_S$, varies linearly with the constructed short-lifetime fraction for three different total photon numbers, with a small offset. Squares: estimate from posterior mode. Dots: estimate from posterior mean. Dashed lines: linear fits with slopes $0.9933 \pm 0.0026$, $1.0085 \pm  0.0024$, and $1.0106 \pm 0.0031$, and offsets of $0.002000 \pm  0.0484 \times 10^{-4}$, $ 0.004400 \pm 0.0814\times 10^{-4}$, and $0.009500 \pm 0.1992\times 10^{-4}$ for low, medium, and high intensities respectively (95\% confidence interval). Intensities correspond to data collected at $\approx 1.5 \times10^5$, $1.2 \times 10^6$, and $4.8 \times 10^6$  counts per second, for low, medium, and high intensity respectively. Inset: Data from main figure shown on a log-log scale. (C) Changes in the estimated short-lifetime fraction track the known changes in the short-lifetime fraction. Squares: estimate from posterior mode. Dots: estimate from posterior mean. Dashed lines: Linear fits with slopes of $0.9573 \pm  0.1895$, $1.0013 \pm 0.1445$, and $0.9580 \pm 0.1717$, and offsets of $(0.2332 \pm  0.3712)\times 10^{-5}$, $(0.3215 \pm 0.3088)\times 10^{-5}$, and $(0.4617 \pm 0.4286) \times 10^{-5}$ for low, medium, and high intensities respectively (95\% confidence interval). (D) Sample standard deviations decrease with increasing photon number as $\approx \sqrt{n_{photon}}$ Squares: Posterior standard deviation Dashed line: power law fit to all intensities with exponent $-0.4764 \pm 0.0471$ (95\% confidence interval)}
\label{fig:Figure3}
\end{figure}

We next tested our results in the regime where a relatively large number of photons are collected, but the fraction of photons originating from the short-lifetime component is low.  This regime is relevant in systems where a large number of donor molecules are present, but interactions leading to FRET are relatively rare. In order to test the performance of our algorithm in this regime, fluorescence lifetime measurements were taken of Erythrosin B and Coumarin 153 as representative short- and long-lifetime dyes respectively. Unlike the measurements taken in the low-photon regime, separate fluorescence lifetime measurements were taken for each dye, generating separate master photon histograms (Fig. \ref{fig:Figure3}A). A fixed number of photons could then be numerically sampled from each master histogram in order to create test histograms containing a prescribed fraction of photons originating from the short-lifetime dye, which were then analyzed in order to estimate the short-lifetime fraction. Data was collected at $\approx 1.5 \times10^5$, $1.2 \times 10^6$, and $4.8 \times 10^6$  counts per second, corresponding to low, medium, and high intensity respectively, and histograms from each intensity were analyzed separately.  

Photons were sampled from master curves such that the total number of photons was fixed at $ 5 \times 10^7$, with a prescribed fraction of photons originating from the short-lifetime distribution. This process was repeated 100 times for each condition. Across orders of magnitude, the short-lifetime fraction estimated from our algorithm varies linearly with the prescribed short-lifetime fraction (Fig. \ref{fig:Figure3}B), with linear fits giving slopes of $0.9933 \pm 0.0026$, $1.0085 \pm  0.0024$, and $1.0106 \pm 0.0031$, and offsets of $0.002000 \pm  0.0484 \times 10^{-4}$, $ 0.004400 \pm 0.0814\times 10^{-4}$, and $0.009500 \pm 0.1992\times 10^{-4}$ for low, medium, and high intensities respectively (95\% confidence interval). The estimated short-lifetime fraction differs from the known short-lifetime fraction by a small bias factor, evident by the small positive offsets in the linear fits (Fig. \ref{fig:Figure3}B, Inset). We hypothesize that this offset may be due to a number of factors, including non-monoexponential photon emission from the dyes, slight mischaracterization of the lifetimes or the instrument response function, or an intensity dependence of the FLIM measurement system. While the magnitude of the bias varies with intensity, the magnitude of the bias is relatively small, overestimating the fraction by less than one percent for the highest intensity tested. 

 In many applications, the changes in FRET fraction are more relevant than the actual fraction values themselves. Thus, we next considered the accuracy of measuring changes in the short-lifetime fraction, which were derived from the results in Fig. \ref{fig:Figure3}B by subtracting values adjacent on the short-lifetime fraction axis. While the estimated short-lifetime fractions contain a small bias (Fig \ref{fig:Figure3}B), the bias is largely removed when changes in short-lifetime fraction are considered (Fig. \ref{fig:Figure3}C). Consistent with this removal of bias, fitting linear equations to the estimated changes in short lifetime fraction vs. prescribed short lifetime fraction gives slopes of $0.9573 \pm  0.1895$, $1.0013 \pm 0.1445$, and $0.9580 \pm 0.1717$, and offsets of $(0.2332 \pm  0.3712)\times 10^{-5}$, $(0.3215 \pm 0.3088)\times 10^{-5}$, and $(0.4617 \pm 0.4286) \times 10^{-5}$ for low, medium, and high intensities respectively (95\% confidence interval) (Fig. \ref{fig:Figure3}C). These results demonstrate the accuracy and precision of our method for measuring changes in short-lifetime fraction across many orders of magnitude. For a short-lifetime fraction of $2^{-7}$, the sample standard deviation decays with increasing photon number. Fitting a power law yields an exponent of  $-0.4764 \pm 0.0471$ (95\% confidence interval), consistent with the exponent of $-0.5$ predicted from the central limit theorem and as was the case for the low-photon regime measurements (Fig \ref{fig:Figure2}C).

\section*{Discussion}
Here we presented an extension of previous Bayesian inference approaches to FLIM data analysis that takes into account additional experimental complexities. Using controlled experimental data as a test case, we show that this analysis performs remarkably well in both the low-photon and low-fraction regimes. 

In the low-photon regime, we can estimate the low-lifetime fraction, $f_S$, with a precision of 0.003 and a bias of 0.017 using only 200 photons. At a photon collection rate of $2 \times 10^5$ photons per second, this number of photons corresponds to an acquisition time of only 1 millisecond. As the precision scales as $\propto n_{photon}^{-1/2}$ (Fig \ref{fig:Figure2}C), if one instead wanted a higher precision of 0.001, one could instead collect data for 9 milliseconds. In the low-fraction regime, using $5 \times 10^7$ photons, for a short-lifetime fraction, $f_S$, of 0.0156, we find a precision of 0.000096 and a bias of 0.0046. With an acquisition rate of $1.5 \times 10^6$ photons per second, this corresponds to $\approx$ 33 seconds of acquisition time. As the precision in this regime also scales $\propto n_{photon}^{-1/2}$ (Fig \ref{fig:Figure3}C), if one requires a higher precision of 0.000032, this could be obtained by acquiring data for nine times as long, or 300 seconds. Thus, in both the low-photon and low-fraction regimes, our results show the required number of photons, and hence the acquisition time, necessary to achieve a given level of precision.

One limitation of our implementation is that we evaluate the posterior distribution at equally spaced points. A large parameter space must be searched, and the analysis presented here is relatively slow compared to other parameter searching techniques. For example, when 4 parameters are searched using a Markov chain Monte Carlo approach to stochastically optimize our likelihood, the computation time is reduced by a factor of $\approx$10-20 with no loss of accuracy (Fig. \ref{fig:SFigure4}).

Here we have focused on the use of FLIM to measure changes in FRET, yet it has wider applications, including in metabolic imaging \cite{Bird:2005bb} and in measuring local changes in environment, including pH \cite{Lin:2003gn} as well as oxygen \cite{Anonymous:UjRtYiS3} and Zn$^{2+}$ \cite{Ripoll:2015gb} concentrations. The analysis presented here is general, and should be applicable to FLIM measurements in these other systems as well. 
\section*{Supporting Information}

\subsection*{Removal of Time Bins}\label{missing_time}
A limitation of the Becker \& Hickl system used here is that it cannot record photon arrival times for the entire collection period. In addition, time bins can be artificially removed from analysis to preferentially remove contaminating factors neglected in the analysis. For example, biological samples such as \emph{Xenopus laevis} oocyte extracts can contain endogenous fluorophores with a lifetime significantly shorter than the donor lifetime. Photons collected from these endogenous fluorophores have a large contribution to early time bins and thus removal of these bins preferentially removes their effects.

 While not utilized in the data presented here, removal of time bins can be taken into account in the likelihood function constructed above, Eqn  \ref{final_likelihood}, by weighting the population fractions, $f_i$ by,
 
 $$a = \frac{\sum^{kept\ time\ bins} p_{em,S}(t \in b_i | f_S,\tau_S) }{\sum^{all\ time\ bins} p_{em,S}(t \in b_i |f_S,\tau_S) }$$
 $$b = \frac{\sum^{kept\ time\ bins} p_{em,L}(t \in b_i | f_L,\tau_L) }{\sum^{all\ time\ bins} p_{em,L}(t \in b_i |f_L, \tau_L) }$$
 $$c = \frac{\sum^{kept\ time\ bins} p_B(t \in b_i |  f_B) }{\sum^{all\ time\ bins} p_B(t \in b_i | f_B) }$$
 $$ D = f_S a + f_L  b + f_B c$$
 Such that,
\begin{equation}\begin{aligned}p( t|\theta) = \prod_{i=1}^N \Big[\frac{a}{D}f_S \times p_{em,S}(t \in b_i|f_S,\tau_S) + \frac{b}{D}f_L \times p_{em,L}(t \in b_i |f_L ,\tau_L) +\\  \frac{c}{D}f_B \times p_B(t \in b_i |f_B) \Big]^{P_i} \end{aligned}\end{equation}

\setcounter{figure}{0} 
\renewcommand{\thefigure}{S\arabic{figure}}
\begin{figure}[h*] 
\centering
\includegraphics[scale=0.75]{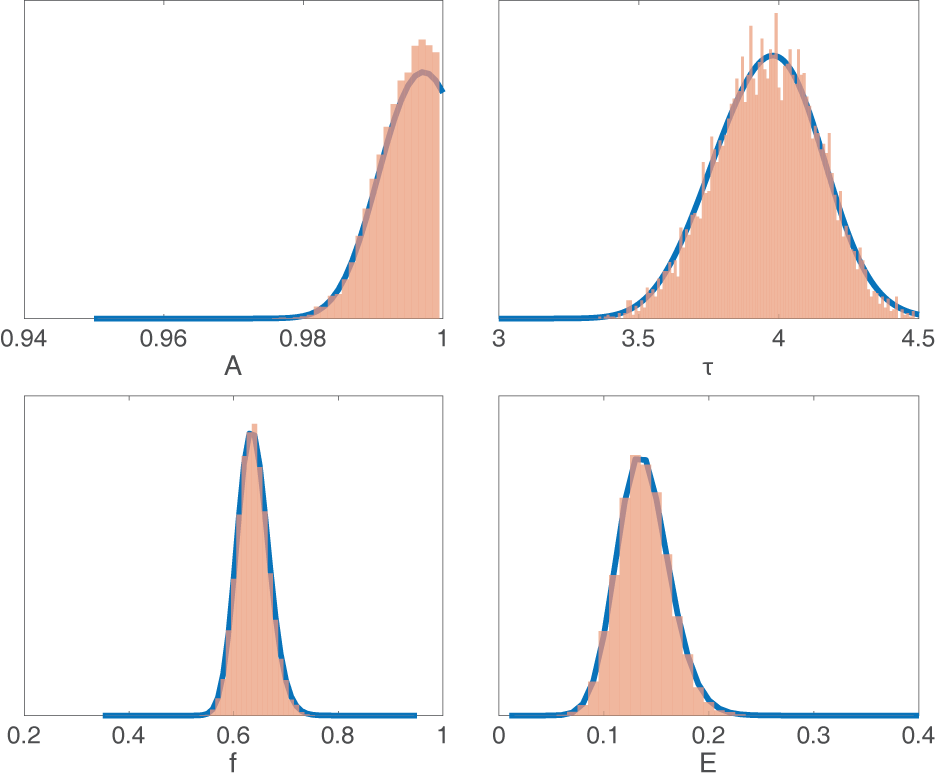}
\caption{{\bf Posterior distributions generated using grid points and stochastic optimization are equivalent.}
Results from Markov chain Monte Carlo (red) and grid points (blue) were generated from the same data set.}
\label{fig:SFigure4}
\end{figure}

\section*{Acknowledgments}
Some computations in this paper were run on the Odyssey cluster supported by the FAS Division of Science, Research Computing Group at Harvard University. The authors would like to thank Jess Crossno and Julia Schwartzman for inspiring us daily. BK was supported by NSF GRFP fellowship DGE1144152. TY would like to thank the Samsung scholarship. This work was supported by National Science Foundation Grants PHY-0847188, PHY-1305254, and DMR-0820484, and
United States--Israel Binational Science Foundation Grant BSF 2009271.

%
%
%

\end{document}